\documentclass[conference]{IEEEtran}
\usepackage[backend=bibtex,style=ieee,natbib=true,sortcites]{biblatex}
\usepackage[most]{tcolorbox}
\usepackage{hyperref}
\definecolor{coolblack}{rgb}{0.0, 0.18, 0.39}
\hypersetup{
unicode=false, 
colorlinks=true, 
linkcolor=coolblack, 
citecolor=coolblack, 
filecolor=coolblack, 
urlcolor=coolblack 
}
\usepackage{booktabs}
\usepackage{tabularx}
\usepackage{enumitem}
\usepackage{url}

\makeatletter
\usepackage{amssymb}
\newcommand{\ygg@basicalert}[2]{\fbox{\bfseries\sffamily\scriptsize#1}{\sf\small$\blacktriangleright$\textit{#2}$\blacktriangleleft$}}
\newcommand{\YANN}[1]{\ygg@basicalert{YANN}{#1}}
\newcommand{\fabio}[1]{\ygg@basicalert{Fabio}{#1}}

\newcommand{\githubpage}{\url{https://github.com/game-dev-database/game-testing}}

\newcommand{\numberOfPapersQuery}{327}
\newcommand{\numberOfPapersTitle}{71}
\newcommand{\numberOfPapersAbstract}{38}
\newcommand{\numberOfPapersFull}{8}
\newcommand{\numberOfPapersSnow}{96}

\begin{document}

\title{A Survey of Video Game Testing}

\author{\IEEEauthorblockN{Cristiano Politowski}
\IEEEauthorblockA{Concordia University\\Montreal, QC, Canada\\
c\_polito@encs.concordia.ca}
\and
\IEEEauthorblockN{Fabio Petrillo}
\IEEEauthorblockA{Université du Québec à Chicoutimi\\
Chicoutimi, QC, Canada\\
fabio@petrillo.com}
\and
\IEEEauthorblockN{Yann-Ga\"el Gu\'{e}h\'{e}neuc}
\IEEEauthorblockA{Concordia University\\Montreal, QC, Canada\\
yann-gael.gueheneuc@concordia.ca}}

\maketitle


\begin{abstract}
Video-game projects are notorious for having day-one bugs, no matter how big their budget or team size. The quality of a game is essential for its success. This quality could be assessed and ensured through testing. However, to the best of our knowledge, little is known about video-game testing. In this paper, we want to understand how game developers perform game testing. We investigate, through a survey, the academic and gray literature to identify and report on existing testing processes and how they could automate them. We found that game developers rely, almost exclusively, upon manual play-testing and the testers' intrinsic knowledge. We conclude that current testing processes fall short because of their lack of automation, which seems to be the natural next step to improve the quality of games while maintaining costs. However, the current game-testing techniques may not generalize to different types of games.
\end{abstract}

\IEEEpeerreviewmaketitle

\section{Introduction}

In the game industry, the first impression is of utmost importance; therefore, only high-quality games can expect to succeed.
In December of 2020, the game ``Cyberpunk 2077''\footnote{\url{https://www.cyberpunk.net/ca/en/}} was released after seven years of development\footnote{\url{https://www.youtube.com/watch?v=cGmWwFpNIHg}}, hundreds of millions of dollars\footnote{\url{http://bit.ly/2M41lHG}}, multiple delays\footnote{\url{http://bit.ly/39M5mbV}}, and high expectations. Upon release, it immediately received strong criticism for its ``buggy state'' and, no long after, it was even removed from the PlayStation Store\footnote{\url{https://www.playstation.com/en-ca/cyberpunk-2077-refunds/}} because it was not delivering the expected experience to its users, causing revenue loss and other collateral effects, like loss of prestige and good-will for the company CD Projekt Red\footnote{\url{https://en.cdprojektred.com/}}. 

Many observers in industry and academia ponder how a company with a reputation for quality games\footnote{The previous game released by the same company, ``The Witcher 3'', was considered a great success:  \url{http://bit.ly/3ipMHqs}.}, like CD Projekt Red, with no shortage of money, time, skill, and experience could release a game with such poor quality. These observers, as well as personnel from the company itself, blame the scope, the management, and the \emph{lack of testing}\footnote{\url{http://bloom.bg/2LNtNhc}} for the poor quality of the game. According to the company CEO, Marcin Iwiński, ``Every change and improvement needed to be tested, and as it turned out, our testing did not show a big part of the issues you experienced while playing the game''\footnote{\url{http://bit.ly/39N8epd}}.

Thus, like with any piece of software, to deliver high-quality games, game developers must test their games rigorously during their development. In traditional software development, tests are considered essential (unit, component, integration, or end-to-end tests), and so is their automation \cite{Spinellis2017}. Yet, to the best of our knowledge, there exists no comprehensive and systematic report on how game developers test (or not) their games. Thus, we ask the question: ``What is the testing process in game development?''

To answer this question, we study the field of testing in game development by surveying the processes, techniques, gaps, concerns, and point-of-views using two different sources: (1) the \textit{academic literature} (journal and conference papers, books, and theses) and (2) the \textit{gray literature} (presentations in well-known conferences, blog posts from game developers, and video-game postmortems).

We report that manual game-play testing, a form of end-to-end testing also called play-testing, is the primary testing technique used by game developers. Automate play-testing sessions is difficult, which lead to developers hiring specialised testers to manually play-test their games, resulting in a game industry that relies almost exclusively upon manual labour and human expertise to test their games.

We conclude that the automation of video-game testing requires research and tools that game developers can integrate into their testing processes. Therefore, allowing them to create more systematic, repeatable, and reliable games, while game testers focus on player-centered aspects of the games, like game play or aesthetics.

The paper is structured as follows:  
Section~\ref{sec:method} describes the method we used.
Section~\ref{sec:results-academic} shows the results of our analysis of the academic literature.
Section~\ref{sec:results-gray} shows the results of our study of the gray literature.
Section~\ref{sec:discussion} discusses our results.
Section~\ref{sec:threats} describes threats to validity of our study. Section~\ref{sec:conclusion} concludes the paper with future work.


\section{Survey Method}
\label{sec:method}

Our approach is based on categories of reliability of the literature \cite{Adams2017}, with \emph{white literature}, peer-reviewed papers, is the most reliable, followed by three gray literature tiers: first-tier, books, books chapters, government reports; \emph{second tier}, annual reports, news articles, videos, presentations, Wiki articles; and, \emph{third tier}, blogs, emails, tweets, letters.

We considered: (1) \textit{academic literature}, journal, conference, and workshop papers, books and theses, and (2) \textit{gray literature}, presentations in well-known conferences, blog posts from game developers, and video-game postmortems. We did not observe any duplication or extension (journal papers extending explicitly conference papers with new material). We performed the search for academic and gray literature on 2019/06/21.

\subsection{Academic Literature} \label{sec:method-academic}


We used two sources of academic documents: Scopus\footnote{\url{www.scopus.com}} and Engineering Village\footnote{\url{www.engineeringvillage.com}}.
The query was: (test OR testing OR verification OR validation OR qa OR "quality assurance" OR debugging OR prototyping) AND (game OR video-game OR "video game" OR "digital game") AND NOT (gamification OR "serious games" OR education OR teaching) AND LIMIT-TO (LANGUAGE,  "English"). After the inclusion and exclusion criteria, the query yields \textbf{\numberOfPapersQuery{} papers}.

The \emph{inclusion} and \emph{exclusion} selection criteria were:

\begin{itemize}
\item Inclusion criteria:
\begin{itemize}
    \item Papers must be about video game development;
    \item Papers must be about or discuss testing;
\end{itemize}
\item Exclusion criteria:
\begin{itemize}
    \item Papers not written in English;
    \item Papers about gamification;
    \item Papers about serious games;
    \item Papers about using games for educational purposes;
    \item Papers about using games for medical purposes.
\end{itemize}
\end{itemize}

We did not set and use any quality criteria. 

\paragraph{Filtering: Title Reading} We read the title of each of \numberOfPapersQuery{} papers and divided them int \emph{game development} and--or \emph{testing}. After that filtering, we obtained \textbf{\numberOfPapersTitle{} papers}.

\paragraph{Filtering: Abstract Reading} We read the abstracts of the \numberOfPapersTitle~papers and \emph{rejected} or \emph{accepted} them based on our criteria. We kept \textbf{\numberOfPapersAbstract{} papers}.

\paragraph{Coding: Full Reading} We read all \numberOfPapersAbstract~papers, taking notes of the relevant subjects and references according to the grounded theory technique and selected the most relevant papers for the snowballing, \textbf{\numberOfPapersFull{} papers}.

\paragraph{Snowballing} We performed a complete snowballing process, backwards and forward, on the \numberOfPapersFull{} papers, re-checking the title and abstract of each paper again to verify the paper's relevance. The final dataset\footnote{The dataset is available at \githubpage.} included \textbf{\numberOfPapersSnow{} papers}.

\subsection{Gray Literature} \label{sec:method-gray}

\newcommand{\sourceA}{Postmortems of video-game projects}
\newcommand{\sourceB}{Conferences on game development}
\newcommand{\sourceC}{Media specialized in game development}

For the gray literature, we used the same inclusion and exclusion criteria than the academic papers. We used three main sources of gray literature:

\begin{itemize}
    \item \sourceA
    \begin{itemize}
    	\item We used a database of postmortems \cite{Politowski2020dataset} and kept only the ones related to testing. We analyzed each one of them and extracted the root cause of the problems.
    \end{itemize}
    \item \sourceB
    \begin{itemize}
    	\item We searched for talks about testing in game-developer conferences. We read/watched presentations and summarized the points related to game testing.
	\end{itemize}
    \item \sourceC
    \begin{itemize}
    	\item We searched for articles about testing on Web sites and blogs about game development. We selected the most relevant and summarized the main points.
    \end{itemize}
\end{itemize}

We also searched specialized forums. We searched StackExchange\footnote{\url{gamedev.stackexchange.com}} for questions with tags ``testing''\footnote{\url{https://gamedev.stackexchange.com/questions/tagged/testing}} and found 79 questions but without relevant answers, about  hardware compatibility (11 questions), game-play balancing (10 questions), and network testing (8 questions).

Finally, we queried Google Search: \texttt{test* AND game AND development -job -course -company} and used the five first result pages. We found 100 Web sites (blogs, forums, companies, papers, and courses) but they were irrelevant and--or did not add new information.


\newcommand{\rqA}{What are the \textbf{particularities} of game testing?}
\newcommand{\rqB}{What are the \textbf{techniques} used for game testing?}
\newcommand{\rqC}{What are today's \textbf{concerns} of the field of game testing?}
\newcommand{\rqD}{What are future \textbf{challenges} on game testing?}

\section{Results - Academic Literature} \label{sec:results-academic}

\subsection{Game testing and its particularities}

The game industry has different characteristics compared to traditional software development.
It is a junction between design, business and software engineering \cite{Hyrynsalmi2018}.
In traditional software development, the product's objective is to provide a service while the game's goal is the entertainment \cite{Kasurinen2017}. Traditional software products, called ``productivity applications'', are designed to accomplish one or more tasks as necessary elements of a non-entertainment endeavour (e.g. business endeavours) \cite{Callele2015}.
On the other hand, a game is more than software as it must enthrall the user with its mechanics, despite its foundation relying on a well-written software \cite{Kanode2009}.
Finally, game software aims to provide an experience rather than say productivity, and these cause divergences on the adopted practices \cite{OsborneOHagan2014}.

Game development consists of iteration of experimentation while embracing the changes as the key for a good game design is the constant experimentation of new features instead of stick with a set of requirements \cite{Lewis2011}. Since game development requirements are continually changing (volatile and mutable \cite{Santos2018}), developers prefer to embrace the change, as said by the developer: ``Why should we bother with strict decisions when we can make a prototype in two hours to test things out?'' \cite{Kasurinen2013}. Therefore, game development emphasizes evaluating user experiences and using the feedback to drive design iterations \cite{OsborneOHagan2014}, although some authors report a balance between planned activities and improvised ones \cite{Stacey2009}.

Some games have a particular way of profiting and, therefore, different ways to manage their development life-cycle. For example, mobile games are not designed to be sold once, but to keep changing (balancing the gameplay) and evolving (adding new features) constantly depending on the audience needs and behaviour \cite{Mozgovoy2018}. Regarding the quality, software quality can be determined by how well the product performs its intended functions. For game software, this includes the quality of the player's experience plus how well the game features are implemented \cite{Schultz2005}.


\subsubsection{Game testers}

Gameplay testers have the heavy-duty of finding bugs and any other abnormality in the game. Game testers should test game quality by verifying gameplay, logical consistency, observability, progressive thinking, reasoning ability, and exhaustively testing features, game strategy, and functionality \cite{Aleem2016a}. Therefore, testers should understand the principles and the characteristics behind games and especially understand the game development context \cite{Santos2018}.

Despite their usefulness, testers are often treated as less important compared to the developers and game designers \cite{SchreierKotaku2017, Kasurinen2017}.
Often, they work in non-desirable conditions \cite{SchreierKotaku2017}, and the setup of game testing is cumbersome for the developer. Testing changes will take longer as the programmer needs to set up the proper conditions within the game world to exercise that code path \cite{Blow2004}.
Finding bugs is repetitive, time-consuming and challenging for human testers, which could cause testers to overlook defects \cite{Hooper2017}.

\subsubsection{Game testing}

Testing games have some different aspects compared to testing traditional software \cite{Murphy-Hill2014}.
Video games are not considered productivity software; they are tested less intensely on accuracy and more intensely on the overall experience \cite{Shelley2013}. This type of testing is referred to \emph{gameplay testing}, which is the way developers assess the quality of the game \cite{Ramadan2014} and it is preferred in detriment of other techniques \cite{Murphy-Hill2014, Musil2010a} as it gives the team the feedback in the form of ``enjoyment heuristics'' \cite{Ampatzoglou2010}. Accordingly, test management is more ad-hoc in game development, relying heavily on the usability tests with users, not on the documented test plans or predefined test cases \cite{Kasurinen2014a}.

The game development process involves iterations around experiments, and it is expected that the game's code becomes ``throwable'' \cite{Mozgovoy2018}, making the tests obsolete. This practice is not unusual as the game industry is famous for creating new code with each new title. Even when the ``same'' engine or core code is reused for a new game, that engine or code is often modified, repaired, and improved upon \cite{Schultz2005}. These iterations lead to changes in how the game was conceptualized, designed and coded \cite{Stacey2009}. Late changes are allowed and even expected \cite{Kasurinen2014a} with playtesting inspiring new features \cite{Ruonala2016}.


Games have different types of requirements \cite{Callele2015}. Traditional test planning methods cannot satisfy the fun (entertainment and excitement) requirements \cite{Mozgovoy2018}. Also, measures related to fun, entertainment, gameplay and other user experience aspects could be too complex to plan and evaluate software testing \cite{Santos2018}.
The main difference between testing a game and regular software lies in the game requirements and the scope definitions of the project \cite{Santos2018}.

Testing sessions are based mostly on black-box testing with human testers \cite{Aleem2016a}. Despite the differences, testing in games is mandatory, as stated by this developer: ``If a tester comes to say that this does not work, there is not fun in it, you really cannot leave that in the game, you have to fix it'' \cite{Kasurinen2013}.
``

Finally, game developers find it difficult to reuse their code, have less pressure on evolving their systems’ architecture, and are less able to overview the requirements of a project correctly. Furthermore, they report more difficulties in performing automated unit tests than non-game developers \cite{Pascarella2018}.
More particularities in game testing are described by \citet{Santos2018}, like requirements analysis, test planning, test design and execution.

\subsubsection{Test automation}

There are significant differences between games and non-games regarding game developers' difficulties to write automated tests \cite{Pascarella2018}.
There is less automation in game testing, for different reasons shown in \autoref{tab:testing-difficulties}.

\begin{table}[ht]
\caption{Difficulties in test automation for games.}
\label{tab:testing-difficulties}
\begin{tabularx}{\linewidth}{@{}lX@{}}
\toprule
 Difficulty & Description \\ \midrule
 Coupling & It is hard to write the automation given the coupling between the user interface (UI) and game mechanics \cite{Murphy-Hill2014, Ruonala2016, Mozgovoy2018}. \\ \addlinespace
 Scope & Trying to cover all ``paths'' of the game could restrict the game design \cite{Murphy-Hill2014, Santos2018}, sometimes seen as contrary to agility preventing the fast pace of changes \cite{Ruonala2016, Hooper2017}. Developers believe that they cannot cover everything, due the large search space of possible game states \cite{Ruonala2016}, therefore the effort is not worth \cite{Hooper2017}. \\ \addlinespace
 Randomness & The non-determinism on games (multi-threading, distributed computing, artificial intelligence, and randomness) make it hard to hard to assert the correct behavior \cite{Murphy-Hill2014, Hooper2017}. In this case is a ``emergent software'' as its randomness is a feature and players' surprise is desirable \cite{Lewis2011}. You need access to the randomness part of the game, otherwise you may never reproduce the bug properly \cite{Schultz2005}. \\ \addlinespace
 Changes & It is difficult to keep the automation as the game design change too often, even core mechanics \cite{Kasurinen2013a, Santos2018}, making its documentation becoming obsolete too fast \cite{Lewis2011}. Also, the source code is temporary and highly likely to change as development progresses \cite{Hooper2017}. \\ \addlinespace
 Cost & A software engineering is more expensive compared to gameplay testers \cite{Murphy-Hill2014, Hooper2017}. The cost of hiring new developers to write additional lines of code into a game to support automation (as well as writing scripts and other test utilities, interfaces, and so on) can be more than it would cost to pay human testers \cite{Schultz2005}. Also, manual re-test after code changes (new build) are expensive due to reproducibility issues \cite{Hooper2017}. \\ \addlinespace
 Time & Programmers usually don't test their games. They usually don't have time to do so \cite{Schultz2005} or they believe that writing tests takes time away from the actual development process \cite{Hooper2017}. \\ \addlinespace
 Fun-factor & Capturing the ``fun'' with an automated test is not possible \cite{Hooper2017} as game testing is human centric and human behavior is difficult to automate \cite{Santos2018}. Also, it is hard to automate games where many events are happening simultaneously. Therefore, automation is used mainly in simple tests \cite{Santos2018}. \\ \addlinespace
 Code & Automated testing code may not be bug-free, nor be reusable from one game to another, or from one platform to another \cite{Hooper2017}. Game developers have more difficulties than other developers when reusing code \cite{Pascarella2018}. \\
 \bottomrule
 \end{tabularx}
\end{table}

Despite the difficulties in applying automating tests in game development, there are some benefits of doing it, for example, reproducibility, quality, fewer bugs, better test focus, game stability, and fewer human testers \cite{Hooper2017}.


\subsection{Game testing concerns}

As the workload for gameplay testing is tiresome \cite{SchreierKotaku2017}, the testers often got motion sickness by constant re-playing the games \cite{Murphy-Hill2014}. In this sense, automation can reduce the burden on both developers and human testers, however it also can introduce cost to game production, and it required programmers with a specialized focus on automated testing \cite{Hooper2017}. At the same time, the lack of test automation in game development ``hurts'' the bug fixing process as it becomes harder to reproduce the steps \cite{Murphy-Hill2014}.

Game developers overlook the SE practices. On testing, this happens because low-level testing is neglected for the sake of gameplay testing \cite{Murphy-Hill2014}.
Regarding automated game testing, few studios use test automation alleging causes like development time, staff size, or little knowledge about these tools \cite{Musil2010a}.
Usability tests and frameworks are also used within game development, but they are tailored specifically for the game domain \cite{Wang2014}.
Even pair and test-first programming, two practices well established in traditional software development, were not largely used by game companies \cite{Koutonen2013}.

In game development, software working according to its functional requirements is not enough; the game must be appealing for the user \cite{Scacchi2015}.
This appeal involves balancing the gameplay, which is unique in game development and a challenge for the developers \cite{Callele2015}.

Creating video games is a complex task. The game industry is on the edge of technical complexity. On top of that, the field of game development is hard to assess considering the restrictions of proprietary code \cite{Lewis2011}.


\subsection{Game testing challenges}


\paragraph{Add game testing sooner to development pipeline} Game studios start testing for \textit{User Experience} of games too late in the development life cycle, sometimes as late as beta, which means that most of the feedback obtained from the tests is unlikely to have an impact on the final game \cite{Paakkanen2014}. Key game concepts should also be testing before release so the reactions of the players can be verified \cite{Washburn2016}. Every aspect of a game should be tested during the development and production phases. The most important aspect of testing for game developers is to integrate testing as part of the production phase to improve efficiency. To ensure the delivery of quality games to the market, developers must consider different testing options during the production phase \cite{Aleem2016a}.

\paragraph{Instrument the game with meaningful logs and create tools to visualize the events} Combine playtesting with play analytics and advanced visualizations (e.g., using synthetic, procedurally generated game worlds to visualize game play data sets and temporal relationships) \cite{Scacchi2015}.

\paragraph{Add automating testing suitable for game development domain} The use of manual testing and the challenges with automated testing in game systems highlight the need for a new set of methodologies to ease the developers’ ability to identify malfunctions and to enable automatic testing activities for games. Studies should investigate new record--replay tools, automated test-data generators, etc. \cite{Pascarella2018}.

\paragraph{Define design patterns for game development} As the GoF patterns \cite{gamma1994design} are specific for OO systems, game development patterns are good practices specific for game genres \cite{Lewis2011}. Aside from game design patterns, which other patterns, more related to source code and more similar to traditional software, could be applied when building a game? How the quality of code in games influence the game's functional requirements?

\paragraph{Create an universal game design language}
The creation of a ``universal game design language'' could allow the game developers reuse/extend a feature from game to game \cite{Lewis2011}.
Can developers interchangeably develop a game regardless of the tools (game engine)?


\section{Results - Gray Literature}
\label{sec:results-gray}


\subsection{\sourceA}
\label{sec:results-gray-pms}

We used a database of postmortem problems provided by a previous work \cite{Politowski2021}. Among all the problems, only 4\% (50 out of 1,035) are about testing. The list of test-related postmortems is available on-line\footnote{\githubpage}, whose problems \autoref{tab:summary-problems-pms} summarizes.

\begin{table}[ht]
\caption{Summary of the testing problems found in the postmortems.}
\label{tab:summary-problems-pms}
\begin{tabularx}{\linewidth}{@{}Xr}
\toprule
Problem & N \\ \midrule
Insufficient testing & 22 \\
Process and testing plans issues & 18 \\
Specific project requirements & 13 \\
Feedback & 7 \\
Scope & 6 \\
Reproducibility & 2 \\
Logging & 2 \\
No in-house QA & 2 \\
Combinatorial explosion & 1 \\
Bug fixing & 1 \\ \bottomrule
\end{tabularx}
\end{table}

\textit{Insufficient testing} is the most common problem quoted by game developers. Eight postmortems cite lack of test as a problem (PM\#4,  PM\#19,  PM\#33,  PM\#35,  PM\#38,  PM\#43,  PM\#45,  PM\#49). Five mention that more testing was need in the early phases of the development (PM\#5, PM\#6, PM\#21, PM\#22, PM\#39) while two others in later phases (PM\#7, PM\#11). Lack of unit testing (PM\#17) and regression testing (PM\#18) were also cited as well as beta testing (PM\#39), most of it because of time constraints (PM\#9, PM\#24, PM\#26), and play-testing in general (PM\#34).
Finally, developers also mentioned problems regarding testing-tools setup, like continuous integration, testing systems, and automation (PM\#6, PM\#28).

\textit{Lack of unit testing} and wrong decisions for the testing plan are well exemplified by the following quote:
\begin{quote}
\textit{``In the end, we slipped to the side of not having as many unit tests as we would have liked. (...) By the time we noticed those problems and wanted to start adding tests, it was too late because some of that code relied on non-unit test friendly APIs like UIKit or Box2d. (...) We even shipped with a few off edge cases that we knew were buggy but we didn't dare fix weeks before submission.''} -- Postmortem PM\#17
\end{quote}

The benefits of automated tests are illustrated by:
\begin{quote}
\textit{``With automated tests QA could have focused on other tasks and we would faster know if a certain map or feature were broken when running the automated tests'' while ``Without automated testing, every bug fix had a greater potential of introducing hidden bugs into the game, and this cost us a lot of time when regression issues weren't identified until long after they were introduced.''} -- Postmortems PM\#6 and  PM\#28
\end{quote}

Specific projects' requirements were heavily stressed by game developers. Support for multiple platforms and its difficulties were the requirements mentioned the most (PM\#5, PM\#12, PM\#18, PM\#30, PM\#31, PM\#36). Testing multiplayer and--or online games is difficult for developers (PM\#2, PM\#46, PM\#47, PM\#49, PM\#50). Other issues related to outdated video drivers (PM\#30) and stress test (PM\#42).

Game developers also cited problems with test plans and processes. Game developers complained that QA can block a development team (PM\#3,  PM\#40). The test strategies were considered problems in many games (PM\#1, PM\#2, PM\#19, PM\#29, PM\#37, PM\#45). Other problems were the overestimation/underestimation of the time and effort necessary to test (PM\#2, PM\#17, PM\#27, PM\#41, PM\#44, PM\#47, PM\#49, PM\#50). Estimation issues regarding the infrastructure, mostly hardware concerns, were also mentioned by game developers (PM\#6, PM\#8, PM\#20).

Problems with poor feedback and testing sessions were mentioned for several game projects (PM\#9, PM\#11, PM\#16, PM\#22, PM\#30, PM\#39, PM\#45), e.g.:
\begin{quote}
\textit{``Most beta testers are young people who have a lot of time on their hands; that’s great for finding bugs, but it can also be a problem because some of them lack perspective.'' and ``I mean the kind of testing that finds bugs, not suggests new features.'' Postmortems PM\#12 and \# PM\#46}
\end{quote}

Some game projects have too large a scope to be properly tested. Scope is an issue also cited by game developers (PM\#6, PM\#14, PM\#23, PM\#26, PM\#38, PM\#41)

Other problems related to the reproducibility of the bugs (PM\#13,  PM\#25) and bug-fixing (PM\#3), logging problems (PM\#25,  PM\#30),
lack of in-house QA team (PM\#28, PM\#40), and the complexity of game mechanics (PM\#15).

The reproducibility problem is exemplified by:
\begin{quote}
\textit{``The arena bug was subtle, although the results were not, and was not easy to reproduce. We were dimly aware that the bug existed, but its frequency was unclear. Because of the difficulty in reproducing it, the scope of testing the rest of the game and the small size of the QA team, the bug was not properly identified.''} -- Postmortem PM\#13
\end{quote}

The problems with game scope, game mechanics, and insufficient testing are well exemplified by the following quote:
\begin{quote}
\textit{``[the game] had 100 items and five playable characters. 70 percent of the items in [the character] stack, and all the item abilities will affect [him] in some way, so there were so many variables to keep track of that all the testing in the world couldn't have prepared us for launch. (...) In order to fully test all the variables we had in place, it would have taken hundreds of testers several days of extensive play time to fully debug this [game] -- there were bugs that actually took 100,000+ people four weeks to find due to how buried and rare some of them were.''} -- Postmortem PM\#15
\end{quote}



\subsection{\sourceB}
\label{sec:results-gray-talks}

We searched for talks in conferences specialized in game development like GDC\footnote{\url{https://www.gdconf.com/}} and Digital Dragons\footnote{\url{http://digitaldragons.pl/}}. \autoref{tab:gl-conferences} shows the most relevant talks about game testing.

\newcounter{rowcountergl}[table]
\renewcommand{\therowcountergl}{CASE-\arabic{rowcountergl}}
\newcolumntype{N}{>{\refstepcounter{rowcountergl}\therowcountergl}l}
\AtBeginEnvironment{tabular}{\setcounter{rowcountergl}{0}}
\begin{table}[ht]
\centering
\caption{\sourceB.}
\label{tab:gl-conferences}
\begin{tabularx}{\linewidth}{@{}NXlll@{}}
\toprule
\multicolumn{1}{l}{}  & Title & Url \\ \midrule
\label{case-1} & Smart Bots for Better Games: Reinforcement Learning in Production & \url{https://bit.ly/2Hi5lyJ} \\
\label{case-2} & Automated Testing and Profiling for 'Call of Duty' & \url{https://bit.ly/37hf3w1} \\
\label{case-3} & Automated Testing and Instant Replays in Retro City Rampage & \url{https://bit.ly/2vuuSlS} \\
\label{case-4} & It's Raining New Content: Successful Rapid Test Iterations & \url{https://bit.ly/31KNk5I} \\
\label{case-5} & Automated Testing of Gameplay Features in Sea of Thieves & \url{https://bit.ly/2ONBFh8} \\ \bottomrule
\end{tabularx}
\end{table}

\subsubsection{[\ref{case-1}] Smart Bots for Better Games: Reinforcement Learning in Production}

Ubisoft\footnote{\url{https://ubisoft.com}} showcased their use of Reinforcement Learning (RL) in production to test games with bots that can evolve while playing. RL has been applied/studied mainly in two areas of game development: improving AI behavior and creating test assistants. The former got attention when the OpenAI team trained bots on Dota 2 (OpenAI Five)\footnote{\url{https://openai.com/projects/five/}}. The latter approach uses agents to test games, which has been used by Ubisoft in Triple-A games like Far Cry: New Dawn (2019), which has an open world, and Rainbow Six Siege (2015), which is multiplayer.

The author also described three techniques to train agents using RL: learning from pixels \cite{Mnih2015, Harmer2018}, from states \cite{Horgan2018, Zhao2019}, and from simulation \cite{Lowe2017}.

\subsubsection{[\ref{case-2}] Automated Testing and Profiling for Call of Duty} \label{case-2-cod}

Electronic Arts (EA) is a large company that develops and publishes games. EA has many studios working on different games. The talk was about the franchise Call of Duty, which receives a new iteration almost every year since 2003.

The team developed a tool called Compass, designed to keep track of builds/testing. The team workflow with Compass consisted of five different checks during the development of the game: continuous integration (CI) module, ``all maps'' testing, nightly tests, and maintenance. The tool work with the help of game-play testing. For example, when testers find a place with a low frame-rate, the tool record this location, adds more characters and collects metrics automatically.


\subsubsection{[\ref{case-3}] Automated Testing and Instant Replays in Retro City Rampage} \label{case-3-rcr}

The author used automated testing techniques to test his indie game, with a small scope. His solution was to record the inputs, in log files, and have the engine re-play them. It allowed him to track down bugs using a simple diff tool on the output files. Advantages of this input--record approach are: automated QA, easy to deal with multiple platforms, and easy to narrow down game-play bugs.


The author also suggested that the game code should be deterministic; the same inputs should yield the same outputs. Otherwise, the input--record approach will deal with different outputs every time, defeating its purpose. Aside from building a deterministic game engine, other things to consider are: initializing all variables, creating own (controllable) random method, and splitting each instance that controls different modules of the game. Finally, especially for multi-platform games, floating-point computations may differ, which might yield different behaviors.

\subsubsection{[\ref{case-4}] It's Raining New Content: Successful Rapid Test Iterations}

Riot is the company developing League of Legends, which had 11 million daily players and 5 million concurrent players at the time of the talk. Thus, one bug, even a small glitch, could affect millions. Thus, they created a process with which testing is carried carefully in different steps to detect/prevent bugs.

Their process consists in daily play-tests with continuous delivery that uses days instead of weeks. They also use automated tests for performance and integration testing, with ad-hoc scripts to test the game loop and front end.


\subsubsection{[\ref{case-5}] Automated Testing of Gameplay Features in Sea of Thieves} \label{case-5-sot}

Rare is the developer of Sea of Thieves, using Unreal Engine. They used the engine features to add tests cases at different levels of abstraction. Rare has more than 23,000 tests and 100,000 asset audit checks. With this automated test process, they managed to keep a constant number of bugs and avoid working overtime. They divided the test into:

\begin{itemize}
\item Actor testing (70\%): This type of tests uses the Actor class of Unreal Engine to have lighter tests compared to integration tests, which have heavy dependencies and do not scale well;
\item Unit testing (23\%): This is the most basic test unit;
\item Map testing (5\%): These tests load portions of the game and check a small action, like the interaction between some actors and objects;
\item Screenshot testing (1\%): This type of tests is used to check the rendering output;
\item Performance testing (less than 1\%): These tests monitor the frame-rate of the game and eventual bottlenecks.
\end{itemize}


\subsection{\sourceC}
\label{sec:results-gray-blogs}

Gamasutra is a portal on everything related to video games and game development. We use Google Search to search Gamasutra for anything related to game and testing since 2010: \texttt{test* AND game site:gamasutra.com after:2010}. From the results, we retrieved the first 50 articles, excluded non-relevant articles from their titles and their contents, and retained the four articles in \autoref{tab:gamasutra-articles}.

\newcounter{rowcountermedia}[table]
\renewcommand{\therowcountermedia}{MEDIA-\arabic{rowcountermedia}}
\newcolumntype{N}{>{\refstepcounter{rowcountermedia}\therowcountermedia}l   }
\AtBeginEnvironment{tabular}{\setcounter{rowcountermedia}{0}}

\begin{table}[ht]
\caption{Gamasutra articles}
\label{tab:gamasutra-articles}
\begin{tabularx}{\linewidth}{@{}NXl@{}}
\toprule
\multicolumn{1}{l}{\#} & Title & URL \\ \midrule
\label{media-1} & Testing for Game Development & \url{https://ubm.io/35jhuNl} \\
\label{media-2} & Differences between Software Testing and Game Testing & \url{https://ubm.io/37ygEOj} \\
\label{media-3} & Unit testing in video games & \url{https://ubm.io/2ZOtZPR} \\
\label{media-4} & Automated Testing: Building A Flexible Game Solver & \url{https://ubm.io/2SWY2U2} \\
\bottomrule
\end{tabularx}
\end{table}

\subsubsection{[\ref{media-1}] Testing for Game Development}
In this article, the author argues that video games, aside of being big and complex, are also software with emergent behavior\footnote{``Behavior of a system that is not explicitly described by the behavior of the components of the system, and is therefore unexpected to a designer or observer'' (\url{http://wiki.c2.com/?EmergentBehavior}).}, i.e., unpredictable.
Moreover, the complex game logic and the unpredictability of the users make it difficult to guarantee that games will ``react gracefully''.

Aside of these difficulties, testing games benefits developers, e.g., by generating knowledge about the games. Testing is important because a game with bugs is not fun to play. Yet, automated testing is hard to achieve and does not replace manual testing, but removes repetitive (regression) testing.

The author also states that TDD, a common way to develop traditional software, might not be suitable when prototyping games due to the constant changes in the game design. Finally, the addition of new features should take precedence over bug fixing, i.e., developers should not spend time and effort on an game that may not even be fun to play.

\subsubsection{[\ref{media-2}] Differences between Software Testing and Game Testing}

The author discusses the differences between testing games and traditional software. He argues that black box testing and automation is the same for software and game. However, he lists some particularities of game testing:

\begin{itemize}
\item User-Experience (UX) testing is not the same as fun-factor testing;
\item Balance testing needs a vast knowledge of the game design and the target audience;
\item Testing the game world or some level can be automated in unique ways, such as having bots moving randomly through the game world to see if they get stuck;
\item AI testing requires that the testers understand what triggers the behavior of different types of non-playable characters (NPC) and how these triggers can be confused by different parameters;
\item Multiplayer/Network testing requires many players simultaneously in the game world, with computer-controlled opponents and different game servers.
\end{itemize}

\subsubsection{[\ref{media-3}] Unit testing in video games}

In this article, the author discusses some concerns related to unit tests in game projects. He stated that he worked on Triple-A projects that lasted more than 3 year, with 20 developers, and without a single unit test. Most developers were highly specialized, working only on a tiny fraction of the whole project; for example, terrain rendering, network, UI, game-play, etc.

\subsubsection{[\ref{media-4}] Automated Testing: Building A Flexible Game Solver}

This article is one more example of an ad-hoc solution to automate game testing. They used a script describing the goals of each game level and a brute-force algorithm to reach these goals\footnote{\url{https://www.youtube.com/watch?v=yXXTJ_EKOIQ}}. Although this solution is well-suited for their game, there are problems with hard-coded test scripts, for examples:

\begin{itemize}
\item Linear testing: actions are always in the same order;
\item Cross quests: a test script cannot take into account variations or optional side quests;
\item Up to date: if developers ``forget'' to report modifications to the adventure, the scripts may fail;
\item Dependent upon development completion: it is difficult to test until the complete adventure is implemented;
\item Logic: a change to a quest in one chapter may impact another part of the game.
\end{itemize}

\newcommand{\fin}{Finding~}
\newcommand{\obs}{Observations~}
\newcommand{\imp}{Implications~}
\newcommand{\source}{Source~}

\section{Discussion} \label{sec:discussion}



\subsection{\fin \#1: The testing strategies should take into account the particularities of game projects wrt.\ traditional software} \label{fin-1}

\paragraph*{\obs - Game development process has particularities} 
The majority of the researchers agree that game development has particularities compared to traditional software development, mostly because of the final product goal: software as a ``productivity'' product and video game as an ``entertainement'' product. These differences reflect on the way the process of developing a game is conducted, mainly with requirements and addition of features along the development. 

\paragraph*{\imp - New game testing methods are necessary}
These particularities of game development should be taken in consideration when devising a testing strategy for a game project. Applying traditional testing strategies, e.g. high emphasis on unit testing using Test Driven Development (TDD), to a game project might not bring the expected benefits. Yet, new testing strategies 
should be the results of the convergence of well-known software testing strategies, like unit testing, continuous integration (CI), and TDD. Finally, these particularities do not justify using only manual testing.



\subsection{\fin \#2: Game testers should work alongside software testers to complement one another skills} \label{fin-2}

\paragraph*{\obs - Game testers are not software testers} 
Game testers are mostly game-play testers, responsible for searching bugs and assessing the game experience, i.e., the fun factor. Although they are usually not considered part of the development team and have poor working conditions, their feedback is fundamental to the success of games.

\paragraph*{\imp - Game and software testers must work together as a team}

In game development, testers are professionals with specific skill sets related to game testing, and they are not engineers in general. Therefore, they write, automate, or script source code-related tests. They deal with games as black-boxes. Thus, the test team should include engineers who can automate tests.


Testing in game development is mostly\footnote{Especially with independent developers, teams sometimes cannot afford a separated QA team and must themselves test the games.} done by a QA team. This professionals assess games as black boxes, playing the role of users (players), stressing the games as intended for their users. Game developers deal with coding tasks, like game-engine development. 
Drawing a parallel with traditional software development, game developers should be the responsible for creating and executing tests. 
Thus, testing in game projects should include game developers with knowledge of the source code and game testers with knowledge of the games. 


As with traditional software development, studios might ignore testing for reasons like scope, budget, and time. In these cases, coverage of the games might not be a good metric to follow. 
Still, game developers should pursue ways to automatize the testing process, but also provide ways to aid the game testers in their task.

\subsection{\fin \#3: Automation in game development is overlooked, as it relies on manual human testers} \label{fin-4}

\paragraph*{\obs - Automation in video-game testing is not common}
Automation is often overlooked because of coupling, scope, randomness, cost, requirements, time, domain restrictions, non-reusable code. Yet, automation has many benefits: reproducibility, quality, bug reduction, better test focus, game stability.

\paragraph*{\imp - Lack of automation demands more (from) human testers}
In game development, the concern (and budget) mostly focuses on play-testing the game. 
Thanks deep-learning algorithms, there are initiatives to train agents and play test the games. It is still too early for proper adoption (i.e., using a standard solution), but it is indeed a step forward to mitigate the burden of the repetitive tasks that game testers have to do.




\subsection{\fin \#4: Search for the ``fun-factor'' and ``balancing'' the game are mainly executed by game-play testers} \label{fin-5}

\paragraph*{\obs - Game testing employs exclusive techniques}
According to the academic papers, there are testing techniques exclusive for game development, mostly related to game-play testing and the search for fun or game balancing. However, these new techniques are poorly explained or too abstract, making them hard to implement and automate. Like regression testing or smoke testing, some techniques are known for their use in traditional software development. Still, in game development, the approaches are different, relying more on game testing.

\paragraph*{\imp - Techniques performed only by gameplay testers}
The search for the fun-factor is not the only thing that game testers do. They also use their skills and empirical knowledge to investigate the games intelligently. These techniques are hard to automate.


\subsection{\fin \#5: There is no one-size-fits-all testing process for game projects as the games differ greatly} \label{fin-6}

\paragraph*{\obs - Big studios use \textit{ad-hoc} testing techniques}
Big studios with resources can develop testing processes tailored for their needs, well suited for specific games. Small studios rely on creativity and customized techniques to test their games. The lack of well-defined strategies is evident, even when using of-the-shelves engines like Unreal, developers must devise their own testing strategies.

\paragraph*{\imp - The use of ad-hoc techniques shows that there is no testing standard for game development}
Big studios can build tailored systems that are well integrated into their development pipelines. As game studios struggle to reach the release date, commonalities among different game types should be investigated, so that working strategies could reused/tailored for specific games. 



\subsection{\fin \#6: The acknowledge importance of testing by game studios should open the door for open-source initiatives} \label{fin-7}

\paragraph*{\obs - Game studios are aware of testing importance}
Although not common, there are advanced testing techniques adopted by big companies. All projects described in the gray literature (Section~\ref{sec:results-gray}) implemented a different testing technique, so it can be better applied to its game type. This fact allows us to believe that developers are aware of the importance of discover new techniques to test the game and not rely only on play-testing sessions. We observed that different studies apply ad-hoc game testing strategies, varying according to game genres (FPS, RPG, Sports, etc), game types (2D, 3D), revenue models (game-as-a-service), etc.

\paragraph*{\imp - The testing importance shows that the game industry should invest on open-source initiatives to spread the knowledge and advance the field} 
The game industry should learn from traditional software initiatives and invest in sharing knowledge among game developers, so that the field can grown faster. 







\subsection{Finding \#7: Issues like lack of plan and poor testing coverage call for game testing to be performed early} \label{fin-9}

\paragraph{\obs - Lack of plan and testing coverage are the main problems}
The most common issues gathered from video-game postmortems indicate that game developers do not plan ahead testing and, consequently, test coverage is low. Moreover, the specific requirements of the project are listed as the main cause of difficulties in adding proper testing during the development. Other problems are related to feedback from the testers and project's scope.

\paragraph{\imp - Testing and gameplay testing should not be delayed} 
The non-linearity and randomness of most of the games nowadays, makes it hard to cover essential cases of a game. The scope---the world, the variables, the randomness, the paths---are too much to cover. Developers have concerns regarding the lack of testing. Testers need the game artifact in a ``playable'' version to perform their assessments. However, given the tight deadlines, this playable build is only accessible late in development. Game developers should strive to release incremental builds to game testers.

\section{Threats} \label{sec:threats}

\paragraph{Hard to find trusted information} The most challenging aspect of this study was finding trusted information about video-game development. We saw biased data even in academic papers. Moreover, video-game studios are secretive about their processes and practices. We tried to mitigate this challenge by using three sources of information: academic papers, technical blogs and talks, and postmortem data.

\paragraph{Academic studies diverge greatly} Studies about games and software engineering discuss mainly the differences between systems and games, while AI papers focus on machine-learning models for mastering the game. Few papers show an end-to-end solution for testing games. 

Our choice of using only titles and abstract to classify a paper as relevant may have led us to reject (many) relevant ones. However, these papers should have reappeared during the snowballing process (from \numberOfPapersTitle{} to \numberOfPapersAbstract{} to \numberOfPapersSnow{}).

\paragraph{Absence of a common field} Academic studies are spread across computer science (new algorithms), software engineering (tools and processes), video-game design (game play, aesthetics), and machine learning (training agents). Thus, points of view and vocabularies vary greatly. We kept this study focused on software engineering, but intersections/overlaps and contradictions still exist.

\section{Conclusion}
\label{sec:conclusion}

No matter how large are their budgets or team sizes, video-game projects are notorious for having day-one bugs. Some games can recover, like No Man's Sky that had problems with missing features at launch but recovered its reputation after many updates along the years\footnote{\url{http://bit.ly/3isBBkn}.}, while others are forgotten, like Anthem, which also had bugs at launch but which developers preferred to rewrite it entirely\footnote{\url{http://bit.ly/3ir9PF3}}, finally ending up canceled\footnote{\url{https://www.pcgamer.com/anthem-next-cancelled/}}. 

In this paper, we studied video game projects by investigating game testing in the academic and gray literature. We surveyed processes, techniques, gaps, concerns, and point-of-views to understand how game developers test their projects. 

The main findings are as follows:

\begin{enumerate}
\item Testing strategies must consider the particularities of game projects; unlike traditional software (Section \ref{fin-1});

\item Game testers should work along with software testers to complement each other skills (Section \ref{fin-2});


\item Automation is overlooked; game testing currently relies mostly on human testers (Section \ref{fin-4});

\item The search for the ``fun-factor'' and ``game balance'' is mainly executed by game-play testers (Section \ref{fin-5});

\item There is no one-size-fits-all testing process for game projects as the game types differ greatly (Section \ref{fin-6}); 

\item Game studios acknowledge the importance of testing, opportunity for open-source initiatives (Section \ref{fin-7});


\item Issues like lack of plan and poor testing coverage call for game testing to be performed early (Section \ref{fin-9});

\end{enumerate}

We conclude that automating the video-game testing process is the natural next step. However, developers and researchers lack processes, frameworks, and tools to help them with test automation. It results in \textit{ad-hoc} techniques that are hard to generalize on different game types. Yet, testing is the key to quality games, and automation makes game quality sustainable.

In future work, we want to expand this survey into a complete, systematic mapping of the academic and gray literature, with clear research questions, quality criteria, and including topic (in)dependent classifications, like publication trends. We will compare the contents and trends of the academic and gray literature, in particular separating processes and techniques. We will also focus on practical techniques to automate the game testing, allowing designers, programmers, and game testers to work together.



\printbibliography
\end{document}